# Structure of the Phase in Pure Two-Mode Gaussian States

R. W. Rendell and A. K. Rajagopal
Naval Research Laboratory, Washington DC  20375

The two-mode relative phase associated with Gaussian states plays an important role in quantum information processes in optical, atomic and electronic systems.  In this work, the origin and structure of the two-mode relative phase in pure Gaussian states is studied in terms of its dependences on the quadratures of the modes. This is done by constructing local canonical transformations to an associated two-mode squeezed state.  The results are illustrated by studying the time dependence of the phase under a nonlocal unitary model evolution containing correlations between the modes.  In a more general context, this approach may allow the two-mode phase to be studied in situations sensitive to different physical parameters within experimental configurations relevant to quantum information processing tasks.
PACS numbers: 03.65.UD, 03.67.-a, 42.50.Dv

## I. Introduction

Quantum information processing using continuous quantum variables have emerged as an alternative to discrete-level systems [1] because Gaussian states can be readily produced from reliable sources [2] and controlled experimentally using accessible sets of operations such as beam splitters, phase shifters and squeezers [3] and efficient detection systems.  In particular, the two-mode squeezed vacuum state of photons has played an important role in both quantum optics and quantum information. The production of these states using non-linear optics by passage of light through down-conversion crystals or by the use of beam splitters is well known [4]. The entanglement of two modes of light is directly related to the two-mode squeezing.  Since quantum entanglement is of paramount importance in quantum information and quantum optical computing has recently been shown to be a viable approach towards making a quantum computer [5], a complete understanding of the properties of these states is important. Gaussian states have already been utilized in realizations of quantum key distributions [6], teleportation [7], and electromagnetically induced transparency [8].  Theoretically, only the squeezing strength determines the entanglement of formation whereas the squeezing phase plays a role in determining the Einstein-Podolsky-Rosen (EPR) correlations between the two modes [9].  In the case of one-mode squeezed states of the photon, the squeezing phase rotates the plane of polarization of the photons and is of importance in considerations of the standard quantum limit that controls the measurement of position in recent experiments [10].

The purpose of this paper is to develop an understanding of how the two-mode relative phase in pure Gaussian states depends and is controlled by the quadrature parameters of the modes.  It is known that any bipartite multimode pure Gaussian state can always be decomposed into products of two-mode squeezed states and products of oscillator grounds states by local linear canonical transformations [11-12].  In particular, this means that a general pure two-mode Gaussian state is locally equivalent to a two-mode squeezed state.  Thus the properties of the two-mode relative phase can be most conveniently studied by focusing on the two-mode phase of the associated squeezed state.   In many discussions of two-mode squeezed states, the two-mode phase is set equal to zero for convenience in order to simplify the expressions.  This is sometimes justified because some quantities of interest (e.g. entanglement) are independent of the phase.  Furthermore, the two-mode phase can be adjusted arbitrarily using local unitary rotations and can thus in principle be eliminated from the problem in some cases.

However, the phase can not always be controlled in this way and it appears explicitly in some applications with the result that it reveals important information.  It is well known that various relative and internal phases play key roles in the operation of interferometers and gates [13].

Specific relative phases also appear as by-products of certain experimental situations where they give important information characterizing the system. Such is the case in recent experiments where squeezed vacuum states were injected into an optically dense atomic medium under conditions of electromagnetically induced transparency [8]. In these experiments, the two-mode squeezed state phase between the probe light and a local oscillator influences the quadrature noise of the probe light and the correct value of the phase must be included in order to quantitatively understand the data, as can be seen by Eq.(2) and Fig.4 of Ref.[8]. The physics of this process resulted in a particular two-mode phase which cannot be conveniently eliminated by a rotation. In another context, a specific non-zero two-mode phase played a key role in determining the entanglement of formation of symmetric two-mode Gaussians [14]. It was found that two-mode squeezed states have the special property of having the smallest EPR dispersion of all symmetric Gaussian states with a given value of the entanglement. This property could be exploited to perform the optimization over all decompositions required to find an explicit expression for the entanglement of formation. Although it was not explicitly discussed [14], the particular squeezed states used in the derivation had a two-mode squeezing phase $\varphi=\pi$ and the result can be obtained only with this particular phase. The phase of $\varphi = \pi$ is singled out because, unlike the entanglement, the EPR correlation for two-mode squeezed states generally depends on the phase [9]:

$$F_{sq}(s,\varphi) = 2(\cosh(2s) + \cos(\varphi)\sinh(2s)) \qquad (1)$$

so that values of $\varphi \neq \pi$ lead to an increase in $F_{sq}$. Here, s is the two-mode squeezing strength and the EPR dispersion is defined by $F = <\hat{F}> \equiv Tr(\hat{r}\,\hat{F})$ and $\hat{F} = \Delta^2(\hat{p}_1 + \hat{p}_2) + \Delta^2(\hat{q}_1 - \hat{q}_2)$. From these examples, it is seen that understanding the structure of the relative phases in Gaussian states is of general interest and that is the focus of this paper.

Section II presents an explicit local canonical transformation of a two-mode pure Gaussian to an associated squeezed state with two-mode phase. This transformation is found to be implemented using local squeezing and rotations and solutions which generate two-mode squeezed states are presented. The values of the transformation parameters which yield two-mode squeezed states are determined. The resulting two-mode phase and the two-mode squeezing strength of the associated squeezed state are calculated in terms of the original quadratures. This provides a scheme to investigate the phase in processing tasks. We illustrate this in Sec. III using a nonlocal unitary evolution of a pure state Gaussian which includes correlations between the quadratures. The time-dependent evolution of the canonical transformation parameters and the resulting phase and squeezing strength of the associated squeezed state are studied. Section IV presents concluding remarks.

## II. Properties of the Two-Mode Pure State Phase

Consider two spatially separated quantum modes i=1,2 described by means of field quadratures [15], the amplitude quadrature, $\hat{X}_i = (\hat{a}_i^\dagger + \hat{a}_i)/\sqrt{2}$, and the phase quadrature, $\hat{Y}_i = i(\hat{a}_i^\dagger - \hat{a}_i)/\sqrt{2}$, in analogy to the position, $\hat{q}_i$, and momentum, $\hat{p}_i$ of the original EPR variables expressed in terms of the destruction and creation operators of mode i, $\hat{a}_i, \hat{a}_i^\dagger$, obeying the usual commutation rules (we use units with h =1). The amplitude and phase quadratures, which determine the properties of the optical beams both as to entanglement and polarization correlations, are routinely measured [16-17]. We examine the general normalized two-mode Gaussian wave function in the representation of the amplitude quadrature or, equivalently, the position variables:

$$\Psi(q_1, q_2) = N \exp-\left(\boldsymbol{a}\, q_1^2 + \boldsymbol{b}\, q_2^2 + 2\boldsymbol{g}\, q_1 q_2\right)/2 \tag{2}$$

where $\boldsymbol{a} = \boldsymbol{a}_1 + i\boldsymbol{a}_2$, $\boldsymbol{b} = \boldsymbol{b}_1 + i\boldsymbol{b}_2$, $\boldsymbol{g} = \boldsymbol{g}_1 + i\boldsymbol{g}_2$, $N^2 = \Delta/\pi$, and $\Delta^2 = \boldsymbol{a}_1\boldsymbol{b}_1 - \boldsymbol{g}_1^2 > 0$. Terms linear in q were not included in eq.(2) because these correspond to displacements which do not affect the two-mode phase.

For two modes, the covariance V matrix contains all the necessary information to characterize the state. The 4x4 correlation matrix [18-20] is written in terms of the 2x2 partitioned matrices representing the marginals A, B and the correlations C, $C^T$:

$$V = \begin{pmatrix} A & C \\ C^T & B \end{pmatrix}; \quad A = \begin{pmatrix} \langle \hat{q}_1^2 \rangle & \langle \{\hat{q}_1, \hat{p}_1\} \rangle \\ \langle \{\hat{q}_1, \hat{p}_1\} \rangle & \langle \hat{p}_1^2 \rangle \end{pmatrix}, \quad B = \begin{pmatrix} \langle \hat{q}_2^2 \rangle & \langle \{\hat{q}_2, \hat{p}_2\} \rangle \\ \langle \{\hat{q}_2, \hat{p}_2\} \rangle & \langle \hat{p}_2^2 \rangle \end{pmatrix},$$

$$C = \begin{pmatrix} \langle \hat{q}_1 \hat{q}_2 \rangle & \langle \hat{q}_1 \hat{p}_2 \rangle \\ \langle \hat{p}_1 \hat{q}_2 \rangle & \langle \hat{p}_1 \hat{p}_2 \rangle \end{pmatrix}; \quad C^T = \begin{pmatrix} \langle \hat{q}_1 \hat{q}_2 \rangle & \langle \hat{q}_2 \hat{p}_1 \rangle \\ \langle \hat{p}_2 \hat{q}_1 \rangle & \langle \hat{p}_1 \hat{p}_2 \rangle \end{pmatrix}.$$

(3)

where $\hat{p}_i = -i\partial/\partial q_i$ is the operator conjugate to $\hat{q}_i$, i =1, 2 and $\langle \{\hat{a}\, \hat{b}\} \rangle = \langle \hat{a}\hat{b} + \hat{b}\hat{a} \rangle / 2$. In [9], the elements of eq.(3) have been explicitly calculated in terms of the coefficients of the wavefunction in eq.(2).

From the results of [11-12], eq.(2) is equivalent to a two-mode squeezed state by a local linear canonical transformation. The corresponding elements of the correlation matrix in eq.(3) for the two-mode squeezed state have particularly simple forms:

$$A_{sq} = B_{sq} = \frac{\cosh 2s}{2}\begin{pmatrix} 1 & 0 \\ 0 & 1 \end{pmatrix}, \quad C_{sq} = -\frac{\sinh 2s}{2}\begin{pmatrix} \cos\varphi & \sin\varphi \\ \sin\varphi & -\cos\varphi \end{pmatrix} \tag{4}$$

where φ is the two-mode phase and s is the squeezing strength. The wave-function for the two-mode squeezed state is obtained by applying the two-mode squeezing operator [15], $\hat{S}_{12} = \exp(-z\hat{a}_1^\dagger \hat{a}_2^\dagger + z^*\hat{a}_2 \hat{a}_1)$ with $z = s\exp(i\varphi)$, to the vacuum. This can be expressed either in terms of Fock states: $|\psi_{sq}\rangle = \operatorname{sech}(s) \sum_{n=0}^{\infty} \left[-e^{i\varphi} \tanh(s)\right]^n |n\rangle_A |n\rangle_B$ or directly in terms of the parameters of eq.(2). The latter is obtained by using the squeezed mode annihilation operators, $\hat{S}_{12} \hat{a}_i \hat{S}_{12}^\dagger$, to act on the vacuum, leading to differential equations for a wave function which is a special case of the Gaussian in eq.(2) with α=β, where:

$$\boldsymbol{a} = (1 + \boldsymbol{l}^2)/(1 - \boldsymbol{l}^2), \quad \boldsymbol{g} = -2\boldsymbol{l}/(1 - \boldsymbol{l}^2) \tag{5}$$

and $\boldsymbol{l} = -\tanh(s)\exp(i\varphi)$, $s \geq 0$, such that $\boldsymbol{a}^2 - \boldsymbol{g}^2 = 1$. Writing eq.(5) explicitly in terms of the two-mode squeezed state parameters, we find:

$$\boldsymbol{a}_1 = \boldsymbol{b}_1 = \frac{\cosh 2s}{\Delta(s,\varphi)}, \quad \boldsymbol{a}_2 = \boldsymbol{b}_2 = \frac{\sin\varphi\, \cos\varphi\, \sinh^2 2s}{\Delta(s,\varphi)}$$

$$g_1 = \frac{\cos \mathbf{j} \, \sinh 2s}{\Delta(s,\mathbf{j})}, \quad g_2 = \frac{\sin \mathbf{j} \, \sinh 2s \cosh 2s}{\Delta(s,\mathbf{j})} \tag{6}$$

where $\Delta(s,\mathbf{j}) = 1 + \sin^2 \mathbf{j} \, \sinh^2 2s$.

In order to obtain an explicit solution for the two-mode phase associated with the pure Gaussian state in eq.(2), we must find a local linear canonical transformation which takes Eq.(3) into the form of Eq.(4). The most general linear local canonical transformation, $\hat{S}$, in phase space can be written in terms of single-mode squeezing and rotations of the creation and annihilation operators:

$$\hat{a}'_i = \cosh(r) e^{i\mathbf{q}} \hat{a}_i + \sinh(r) e^{-i\mathbf{y}} \hat{a}_i^\dagger \tag{7}$$

along with its hermitian conjugate $\hat{a}'^\dagger_i$. This can be applied to the effective annihilation operator for the squeezed state $\hat{S}_i (\hat{S}_{12} \hat{a}_i \hat{S}_{12}^\dagger) \hat{S}_i^\dagger$ and then solved for the transformed wavefunction. However, it is more direct to apply it to the correlation matrix involving the canonical coordinates and momenta. If we require that the transformation bring eq.(3) into the squeezed state form, eq.(4), then:

$$A_{sq} = B_{sq} = S_1 A S_1^T, \quad A_{sq} = B_{sq} = S_2 B S_2^T, \quad C_{sq} = S_1 C S_2^T \tag{8}$$

where $S_i$ are 2x2 matrices with unit determinant, and $S_i^T$ is the transposed matrix of $S_i$ in the phase space of the (q,p) variables:

$$S_i = \begin{pmatrix} \cosh r_i \cos \mathbf{q}_i + \sinh r_i \cos \mathbf{y}_i & -(\cosh r_i \sin \mathbf{q}_i + \sinh r_i \sin \mathbf{y}_i) \\ \cosh r_i \sin \mathbf{q}_i - \sinh r_i \sin \mathbf{y}_i & \cosh r_i \cos \mathbf{q}_i - \sinh r_i \cos \mathbf{y}_i \end{pmatrix} \tag{9}$$

Eqs.(8) and (9) will then yield equations which can be solved for the six transformation parameters $(r_1, \mathbf{q}_1, \mathbf{y}_1)$ and $(r_2, \mathbf{q}_2, \mathbf{y}_2)$ that will bring eq.(3) to the two-mode squeezed state form, eq.(4). The general solution to eqs.(8) and (9) is quite involved but the essence of its structure can be understood for the important special case of symmetric Gaussians, where A=B in eq.(3), and we will present explicit details for this case. In this case, the solution can be obtained with a symmetric transformation, $r \equiv r_1 = r_2, \mathbf{q} \equiv \mathbf{q}_1 = \mathbf{q}_2$ and $\mathbf{y}_1 = \mathbf{y}_2 = 0$.

Note that $S_i$ in eq.(9) preserves the determinants of A and C and this can be used to directly determine the value of the squeezing strength, s, without solving for the transformation parameters:

$$\cosh 2s = 2\sqrt{\det A} \tag{10a}$$
$$\sinh 2s = 2\sqrt{\det C} \tag{10b}$$

This can be understood because $\sqrt{\det A}$ represents the Heisenberg uncertainty for the marginals of each of the modes and the entanglement for a pure state is determined by the von Neumann entropy of its marginals and is independent of the two-mode phase [9]:

$$E_{sq}(s) = \left(\frac{\cosh(2s)+1}{2}\right) \ln\left(\frac{\cosh(2s)+1}{2}\right) - \left(\frac{\cosh(2s)-1}{2}\right) \ln\left(\frac{\cosh(2s)-1}{2}\right) \tag{11}$$

The general two-mode pure state Gaussian given in eq.(2) has six parameters, which determine all the ten quantities appearing in the correlation matrix, V. There are four invariants associated with V. The entanglement is determined by detA, depending only on four of the six parameters, while the EPR correlation depends on all six. Theoretical discussions often utilize a local transformation to standard form of V [18-20]:

$$V = \begin{pmatrix} a & 0 & c_1 & 0 \\ 0 & a & 0 & c_2 \\ c_1 & 0 & b & 0 \\ 0 & c_2 & 0 & b \end{pmatrix}$$

which involves one parameter in the case of a pure state, the local invariant $a^2$=detA, where b=a, $c_2$=-$c_1$ and $a^2$-$c_1^2$=1/4. The entanglement does not change under local transformations and depends only on a. For the special case of squeezed states, the standard form rotates the phase to $\varphi = \pi$ while a= cosh(2s)/2. The method of determining the entanglement of formation based on a property $\varphi = \pi$ squeezed states [14] discussed in Sec. I utilized the standard form of V. Other quantities such as the EPR correlation depend not only on the local invariants but also on the correlation terms and can thus change under local transformations. This was seen in eq.(1) for the special case of squeezed states where the EPR correlation depended on the two-mode phase.

The two-mode phase is more subtle than s since it similarly depends not just on the local invariants but also on the correlations between the modes as can be seen by eq.(4). Therefore the transformation parameters, $\theta$ and r, must be found explicitly in order to obtain the phase. Solving eqs.(8) and (9) for the symmetric situation, we find

$$\cos\boldsymbol{q} = \frac{\langle p_1^2 \rangle - \langle q_1^2 \rangle}{\sqrt{(trA)^2 - 4\det A}} \quad (12a)$$

$$\tanh r = \sqrt{\frac{trA - 2\sqrt{\det A}}{trA + 2\sqrt{\det A}}} \quad (12b)$$

This represents a particular local rotation angle $\theta$ and local squeezing strength r which bring eq.(3) to eq.(4). These appear to depend only on the marginal A of the state, however it can also be seen to depend on the correlations in C since detA + detC = ¼ for a symmetric pure Gaussian [9]. The transformation parameters in eq.(12) can then be used to obtain an explicit expression for the structure of the associated two-mode phase in terms of the quadratures in eq.(3):

$$\cos\boldsymbol{j} = \frac{\langle p_1 p_2 \rangle (\cosh(2s) - 2\langle q_1^2 \rangle) - \langle q_1 q_2 \rangle (\cosh(2s) - 2\langle p_1^2 \rangle)}{\sinh(2s)(\cosh(2s) - \langle q_1^2 \rangle - \langle p_1^2 \rangle)} \quad (13)$$

Note that cosh(2s) and sinh(2s) could be eliminated explicitly in terms of detA and detC by using eq.(10). The associated two-mode phase is seen from eq.(13) to be determined by a relative balance between the marginal and correlation terms of the original state, eq.(3), and its associated two-mode squeezed state, eq.(4). This is made explicit by defining the sums and differences, $\Delta A^\pm \equiv A_{sq} \pm A$, $\Delta C^\pm \equiv C_{sq} \pm C$, so that the condition for the phase in eq.(13) can be rewritten as:

$$\frac{\Delta A_{11}^{-}}{\Delta A_{22}^{-}} = \frac{\Delta C_{11}^{-}}{\Delta C_{22}^{-}} \qquad (14a)$$

or equivalently in terms of the off-diagonal elements:

$$\frac{\Delta A_{12}^{+}}{tr\Delta A^{+}} = \frac{\Delta C_{12}^{+}}{tr\Delta C^{+}} \qquad (14b)$$

Eq.(14) quantifies how the marginals and correlations of eqs.(3) and (4) must be balanced in order to define the two-mode squeezed state with phase given by eq.(13). Note that the transformation in eqs.(8) and (9) could be supplemented by unitary rotations and a two-mode squeezed state would still result. Such a rotation by an angle $\phi$ would shift the phase in eq.(13) to $\varphi + 2\phi$. In particular, the phase could be shifted to $\pi$, resulting in a squeezed state in standard form. However, eq.(13) represents the phase associated with the most direct transformation of eq.(3) to a squeezed state and we will use this to the define the associated two-mode phase. This allows us to monitor in detail how specific physical processes manipulate and control the phase. Note that the elements of the correlation matrix can all be determined experimentally. Thus the expressions eqs. (10), (12) and (13) are important in expressing the theoretical quantities (r,θ) and (s,φ) in terms of the measurable elements of the correlation matrix. An example of this is given in the next section.

## III. Nonlocal Evolution Model

To illustrate the results of Sec.II for the structure of the associated two-mode phase, we consider a pure Gaussian which evolves unitarily in time and study the transformation parameters r(t) and θ(t) as well as the associated squeezed state parameters, s(t) and φ(t). Since the explicit results of eqs.(10), (12) and (13) apply to a symmetric state, we require an evolution which maintains the symmetry during its evolution. In addition, it is desirable to use a unitary which is nonlocal in the modes so that the entanglement will also change during the evolution. The simplest nonlocal symmetric unitary evolution corresponds to free center of mass motion of the pair of modes: $\hat{U}(t) = \exp{-it\left((\hat{p}_1 + \hat{p}_2)^2/2\right)}$, where nonlocality arises from correlations between the modes. As an initial state, we begin with a particular two-mode squeezed state $\psi_0$ with parameters $s_0$ and $\varphi_0$ and $\mathbf{l}_0 = -\tanh(s_0)\exp(i\varphi_0)$ as described by eqs. (5) and (6). The time-evolved wave function is obtained by applying the mode operators transformed by both $\hat{S}_{12}$ and $\hat{U}$, $\hat{U}(\hat{S}_{12}\hat{a}_i\hat{S}_{12}^{\dagger})\hat{U}^{\dagger}$, to the vacuum state. This leads to differential equations for the evolved states, $\mathbf{\psi}(t) = \hat{U}(t)\mathbf{\psi}_0$, of the Gaussian form eq.2, but with explicitly time-dependent coefficients:

$$\mathbf{\alpha}(t) = \frac{1 + \mathbf{l}_0^2 + it(1-\mathbf{l}_0^2)}{1 - \mathbf{l}_0^2 + 2it(1-\mathbf{l}_0)^2}, \quad \mathbf{\gamma}(t) = -\frac{2\mathbf{l}_0 + it(1-\mathbf{l}_0^2)}{1 - \mathbf{l}_0^2 + 2it(1-\mathbf{l}_0)^2} \qquad (15)$$

where β(t) = α(t). This is still a symmetric Gaussian, since $\hat{U}(t)$ is symmetric between the two modes, but is no longer in squeezed-state form because $\mathbf{\alpha}(t)^2 - \mathbf{\gamma}(t)^2 = \left[1 + 2it\frac{1-\mathbf{l}_0}{1+\mathbf{l}_0}\right]^{-1}$ is not equal to unity for t > 0. However, we can find an associated two-mode squeezed state at each time by utilizing the transformation parameters eq.(12) with the resulting squeezing strength and phase of the squeezed state defined by eqs.(10) and (13) respectively. Evaluation of eqs.(10), (12), and (13) for this evolution model requires finding the time dependences of the quadratures of the modes and these are determined by the time-dependences of the coordinates and momenta under $\hat{U}(t)$, $q_i(t) = q_i(0) + (p_1(0) + p_2(0))t$ and $p_i(t) = p_i(0)$ (i=1,2). For example:

$$\langle q_1^2 \rangle_t = \langle q_2^2 \rangle_t = \cosh(2s_0)/2 + 2t\,\partial F_{sq}(s_0, \boldsymbol{j}_0)/\partial \boldsymbol{j}_0 + t^2 F_{sq}(s_0, \boldsymbol{j}_0)/2 \qquad (16)$$

where $F_{sq}$ is the EPR dispersion of the initial state given by eq.(1). $\hat{U}(t)$ contains correlations between the modes and this can lead to a decrease in the marginal uncertainties of the particles during the evolution. This occurs for $\partial F_0/\partial \boldsymbol{j}_0 = -\sin(\boldsymbol{j}_0)\sinh(2s_0) < 0$ or $\sin(\varphi_0) > 0$, so that the term linear in time can become negative and $\langle q_i^2 \rangle_t < \langle q_i^2 \rangle_0$ for some time interval. This behavior for the case of free center of mass motion is thus determined by the rate of change of the initial EPR dispersion with phase, $\partial F_0/\partial \boldsymbol{j}_0 < 0$. The quadrature contracts during the time interval $(0, t_m)$ after which it expands. The minimum depends on both phase and squeezing strength and occurs at $t_m = -2(\partial F_{sq}/\partial \boldsymbol{j}_0)/F_{sq}$. This is analogous to the one-mode squeezed state with (p,q) correlations induced by free mass motion, $q_i(t) = q_i(0) + p_i(0)t$, studied by Yuen [21]. This allowed contractive behavior below the standard quantum limit with consequences to precision measurements of position [10]. The evolution studied here has aspects which are an entangled two-mode analog of this.

The time-dependences of the other quadratures in eq.(3) are similarly calculated and used to calculate the transformation parameters $\theta(t)$ and $r(t)$ from eq.(12) which determine the associated squeezed state. These are plotted in Fig.1 for initial squeezed states with $s_0=1$ and $\varphi_0 = \pi/4$, $\pi/3$, $\pi/2$, $2\pi/3$, and $\pi$. The initial value of $\cos\theta$ from eq.(12a) is found to be finite although both numerator and denominator nominally vanish:

$$\cos(\boldsymbol{q}(0)) = \frac{\sinh(2s_0)\sin(\boldsymbol{j}_0)}{\sqrt{\cosh(4s_0) + \cos(\boldsymbol{j}_0)\sinh(4s_0)}} \qquad (17)$$

Thus at t=0, the transformation in eq.(12) applies a sudden finite rotation given by eq.(17) but vanishing single-mode squeezing, r(0)=0, when U(t) is switched on. This rotation shifts the phase of the initial squeezed state to $\boldsymbol{j}_0 + 2\boldsymbol{q}(0)$. The rotations required to maintain an associated squeezed state then gradually increase under the center-of-mass evolution and approach $\boldsymbol{q} = \boldsymbol{p}$ for long times. The corresponding single-mode squeezing strength r shows an initial abrupt rise before going through a peak with a subsequent more gradual increase. The peak is due to the same correlations which led to the contractive variance for $\langle q_i^2 \rangle_t$ and is sensitive to the values of $s_0$ and $\varphi_0$ for similar reasons. For instance, the peak becomes sharper and more pronounced as $s_0$ increases. Note that the peak does not occur for $\varphi_0 = \pi$ since $\partial F_0/\partial \boldsymbol{j}_0 = -\sin(\boldsymbol{j}_0)\sinh(2s_0) = 0$. This is another instance of the special properties of the $\varphi = \pi$ squeezed state which had allowed the entanglement of formation for symmetric Gaussians to be determined [14].

The squeezing strength and phase of the associated squeezed state corresponding to these transformation parameters are calculated from eqs.(10) and (13) with the results shown in Fig.2. The evolution changes the balance between the marginal and correlation terms as depicted in eq.(14) and this requires that the associated phase to adjust its value according to eq.(13) as the system evolves. In particular, a slow initial increase in the phase $\varphi(t)$ is followed by an abrupt shift of the phase and a gradual decrease to zero at long times as seen in Fig.2(a). The abrupt shift of φ(t) occurs just as the corresponding squeezing strength s(t) goes through a minimum in Fig.2b. From eq.(10), the time dependence of s(t) is explicitly calculated to be:

$$\cosh(2s(t))^2 = \cosh^2(2s_0) + 2t\, \partial F_0/\partial \boldsymbol{j}_0 \cosh(2s_0) + t^2(1+(\partial F_0/\partial \boldsymbol{j}_0)^2/4) \quad (18)$$

The minimum is again controlled by $\partial F_0/\partial \boldsymbol{j}_0$. The EPR dispersion is independent of time under the center-of-mass evolution but its value depends on the initial phase, as also indicated in Fig.2b. Since s(t) determines the entanglement by eq.(11), the decrease of s(t) represents a temporary disentangling of the two-mode state. The corresponding evolution of entanglement is displayed in Fig.3. Although the value of the entanglement is independent of the phase by eq.(11), it is seen that the two are correlated by the dynamics of the center-of-mass evolution where the disentanglement is accompanied by a rapid phase shift. The origin of this phase shift can be understood directly from eq.(13) in terms of the dynamics of the quadratures. In this section, we illustrated this using a model correlated evolution, however the two-mode relative phase can be studied for more general controlled time evolutions (e.g. quantum gates), state preparations, and measurements as is clear from the above presentation.

## V. Summary and Conclusions

The focus of this paper on the two-mode relative phase is due to its importance in both quantum optics and quantum information. In particular, we have emphasized the subtle aspects of the underlying structure of the two-mode phase which determines where it manifests itself and where it does not. The properties of the two-mode relative phase were found to be conveniently studied by focusing on the two-mode phase of an associated squeezed state. This was accomplished by an explicit local canonical transformation of a two-mode pure Gaussian to an associated squeezed state with two-mode phase. This transformation was found to be implemented using local squeezing and rotations and solutions which generate two-mode squeezed states. The values of these transformation parameters which yield two-mode squeezed states were determined in terms of the pure state quadratures. These results were illustrated using a nonlocal unitary evolution of a pure state Gaussian which includes correlations between the quadratures. The time-dependent evolution of the canonical transformation parameters and the resulting phase and squeezing strength of the associated squeezed state were studied. In a more general context, this approach may allow the two-mode phase to be studied in situations sensitive to various physical parameters generated within experimental configurations, such as gates or interferometers, relevant to quantum information processing tasks [5].

**Acknowledgement:** The authors are supported in part by the Office of Naval Research. A.K.R. expresses his thanks to Dr. Andrew Williams of Air Force Research Laboratory, Rome, New York as well as Dr. Joseph Lanza and Dr. Kenric Nelson of AFRL/IFEC, Rome, New York for their general support of this research.

# References

1. S.L. Braunstein and P. van Loock, Rev. Mod. Phys. **77**, 513 (2005); *Quantum Squeezing*, edited by P.D. Drummond and Z Ficek (Springer Verlag, New York, 2004); A. Ferrao, S. Oliveres, and M.G.A. Paris, quant-ph/0503237.
2. H.J. Kimble and D.F. Walls, J. Opt. Soc. Am. **B4**, 1450 (1987).
3. M. Reck, A. Zeilinger, H.J. Bernstein, and P. Bertani, Phys. Rev. Lett. **73**, 58 (1994).
4. Y. Shih, Rep. Prog. Phys. **66**, 1009 (2003).
5. W.J. Munro, K. Nemoto, and T.P. Spiller, New J. Phys. **7**, 137 (2005). .J. Munro, K. Nemoto, and T.P. Spiller, S.D. Barrett, P. Kok, and R.G. Beausoleil, quant-ph/0506116; K. Nemoto and W.J. Munro, quant-ph/0507232.
6. F. Grosshans, G. van Assche, J. Wenger, R. Brouri, N.J. Cerf, and P. Grangier, Nature **421**, 238 (2003).
7. T.C. Zhang, K.W. Goh, C.W. Chou, P. Lodahl, and H.J. Kimble, Phys. Rev. **A67**, 033802 (2003).
8. D. Akamatsu, K. Akiba, and M. Kozuma, Phys. Rev. Lett. **92**, 203602 (2004).
9. R.W. Rendell and A.K. Rajagopal, Phys. Rev. **A72**, 012330 (2005).
10. R. Ruskov, K. Schwab, and A.N. Korotkov, cond-mat/0406416; 0411617.
11. A. Botero and B. Reznik, Phys. Rev. **A67**, 052311 (2003).
12. G. Giedke, J. Eisert, J.I. Cirac, and M.B. Plenio, Quantum Inf. & Comp. **3**, 211 (2003).
13. P. Grangier, J.A. Levenson, and J-P Poizat, Nature **396**, 537 (1998); M.G.A. Paris, Phys. Rev. **A59**, 1615 (1999); K.T. Kapale, L.D. Didomenico, H. Lee, P. Kok, and J.P. Dowling, quant-ph/0507150.
14. G. Giedke, M.M. Wolf, O. Kruger, R.F. Werner, and J.I. Cirac, Phys. Rev. Lett. **91**, 107901 (2003).
15. S. M. Barnett and P. M. Radmore, *Methods in Theoretical Quantum Optics,* Clarendon Press, ( Oxford, UK) (1997).
16. N. Korolkova, G. Leuchs, R. Loudon, T.C. Ralph, and C. Silberhorn, Phys. Rev. **A65**, 052306 (2002); W.P. Bowen et al., Phys. Rev. Lett. 90, 043601; O. Glöckl, S. Lorentz, C. Marquardt, J. Heersink, M. Brownnutt, C. Silberhorn, Q. Pan, P. van Loock, N. Korolkova, and B. Leuchs, Phys. Rev. **A68**, 012319 (2003).
17. V. Josse, A. Dantan, A. Bramati, M. Pinard, and E. Giacobino, Phys. Rev. Lett. **92**, 123601 (2004).
18. R. Simon, Phys. Rev. Lett. **84**, 2726 (2000).
19. L-M Duan, G. Giedke, J.I. Cirac, and P. Zoller, Phys. Rev. Lett. **84**, 2722 (2000).
20. G. Giedke, B. Kraus, M. Lewenstein, and J.I. Cirac, Phys. Rev. Lett. **87**, 167904 (2001).
21. H. P. Yuen, Phys. Rev. Lett. **51**, 719 (1983); H.P. Yuen, Chap. 7 in *Quantum Squeezing*, edited by P.D. Drummond and Z. Ficek (Springer Verlag, New York, 2004).

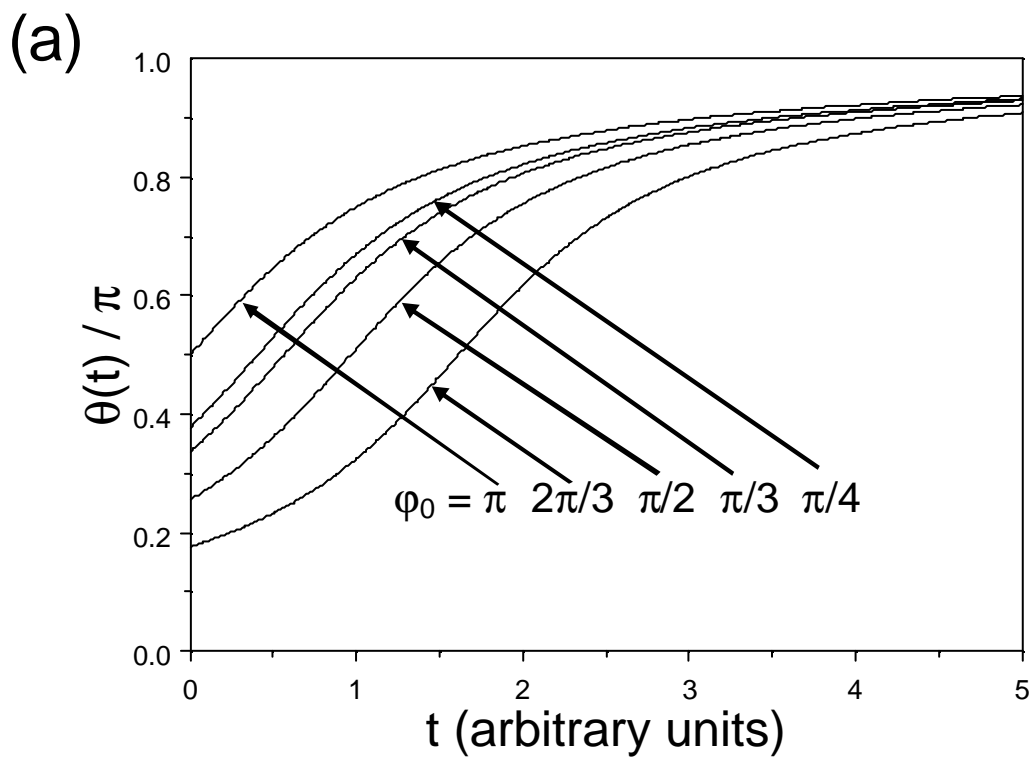

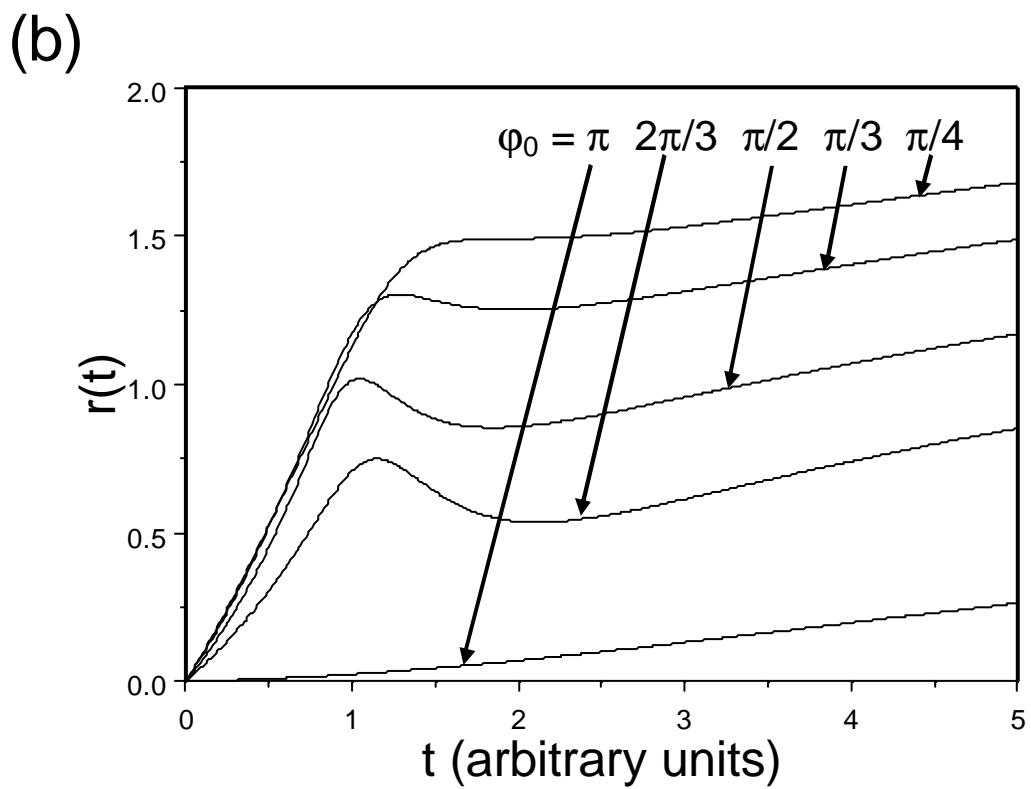

**Figure 1.** Local transform variables, (a) rotation angle θ(t) and (b) single-mode squeezing r(t), bringing the center-of-mass evolution into associated two-mode squeezed states. Initial states have $s_0=1$, $\varphi_0 = \pi, 2\pi/3, \pi/2, \pi/3, \pi/4$.

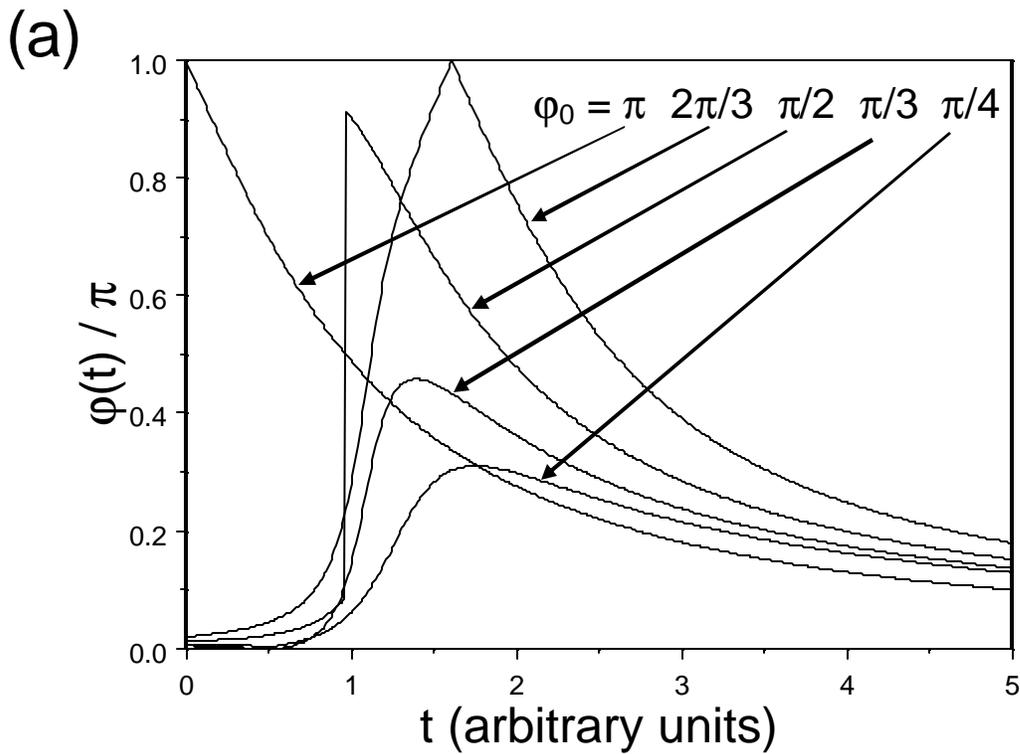

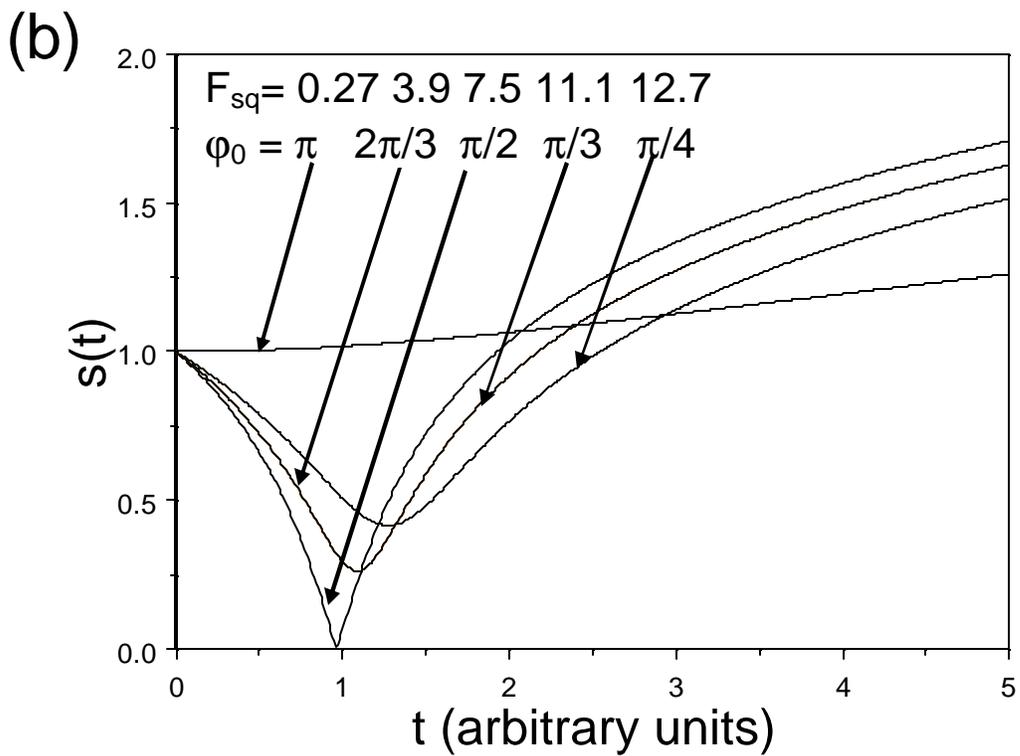

**Figure 2.** (a) Phase, $\varphi(t)$, and (b) squeezing strength, $s(t)$, of the associated two-mode squeezed states for the center of mass evolution. The EPR correlation values from eq.(1) are also given. Parameters are as in Fig.1.

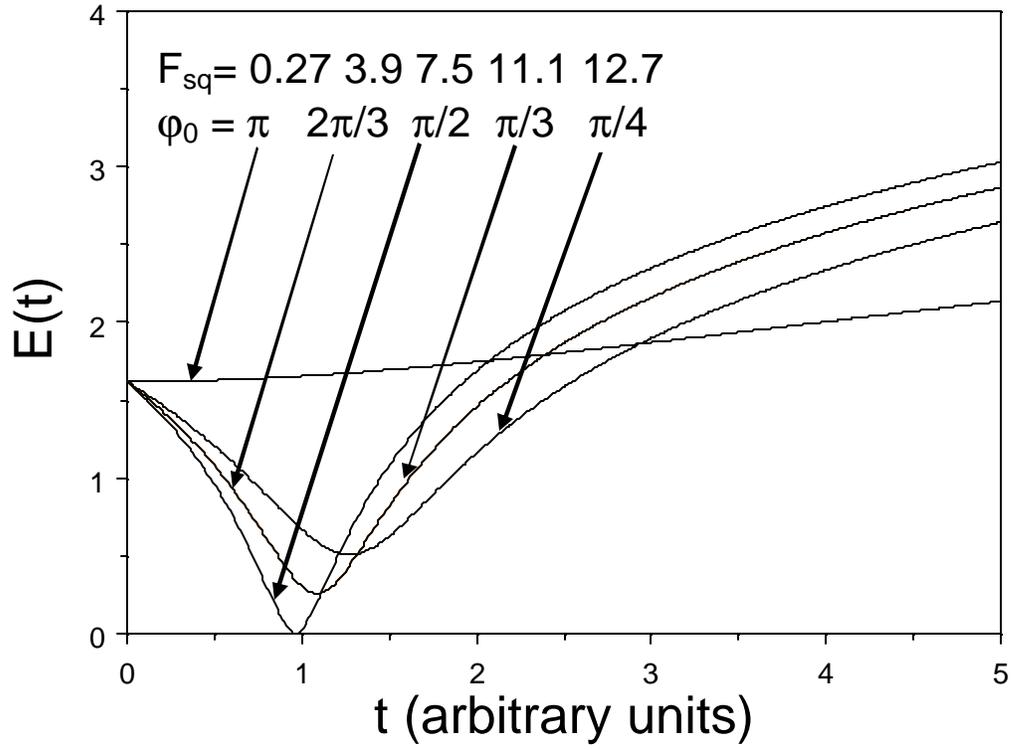

**Figure 3.** Entanglement of the associated two-mode squeezed states from eq.(11) for the center of mass evolution corresponding to Fig.2(b).